\documentclass{article}

\usepackage{prooftree}
\usepackage{latexsym}
\usepackage{epsfig}
\usepackage{amssymb}
\usepackage{amsopn}

\def\NN{\mathbb{N}}

\newcommand{\id}{\operatorname{id}}
\newcommand{\K}[2]{K_{#1} #2}
\newcommand{\J}[2]{J_{#1} #2}
\newcommand{\JT}[3]{J_{#2}^{#1} #3}

\newcommand{\EHAomega}{{\sf E\textup{-}HA}^{\omega}} 
 
\newcommand{\EPAomega}{{\sf E\textup{-}PA}^{\omega}}

\newcommand{\BR}{{\sf BR}} 
\newcommand{\hBR}{{\sf hBR}} 
\newcommand{\EPQ}{{\sf EPQ}}
\newcommand{\TEPS}{T\textup{-}{\sf EPS}}

\newcommand{\DNS}{{\sf DNS}}

\newcommand{\systemT}{{\sf T}}

\newcommand{\forallSt}{\forall^{\hspace{0.2mm}\sf st}} 
\newcommand{\existsSt}{\exists^{\hspace{0.2mm}\sf st}} 

\newcommand{\ba}{\begin{array}} 
\newcommand{\ea}{\end{array}}
\newfont{\bsl}{cmbxsl10 scaled 1095}

\newcommand{\emptyseq}{\langle \, \rangle}

\newcommand{\herb}[1]{(#1)^H}

\newcommand{\power}[1]{{\mathcal P}_{\rm f}(#1)}

\newcommand{\IEP}[1]{{#1}\textup{-}{{\sf EP}}}
\newcommand{\EPS}{{\sf EPS}}

\newcommand{\eqleft}[1]{\begin{itemize} \item[] $#1$ \end{itemize}}

\newcommand{\monApp}[2]{{\rm Ap}(#1)(#2)}
\newcommand{\monAppD}[3]{{\rm Ap}^2(#1)(#2,#3)}

\newtheorem{definition}{Definition}[section]
\newtheorem{theorem}[definition]{Theorem}

\newtheorem{lemma}[definition]{Lemma}

\begin{document} 

%%%%%%%%%%%%%%%%%%%%%%%%%%%%%%%%%%%%%%%%%%%%%%
%%%%%%%%%%%%%%%%%%%%%%%%%%%%%%%%%%%%%%%%%%%%%%
%%%%%%%%%%%%%%%%%%%%%%%%%%%%%%%%%%%%%%%%%%%%%%
\title{The Herbrand Functional Interpretation \\ of the Double Negation Shift}
%%%%%%%%%%%%%%%%%%%%%%%%%%%%%%%%%%%%%%%%%%%%%%
%%%%%%%%%%%%%%%%%%%%%%%%%%%%%%%%%%%%%%%%%%%%%%
%%%%%%%%%%%%%%%%%%%%%%%%%%%%%%%%%%%%%%%%%%%%%%

\author{Mart\'in Escard\'o \and Paulo Oliva}

\date{{\sc Preprint, \today}}

\maketitle

\begin{abstract} This paper considers a generalisation of selection functions over an arbitrary strong monad $T$, as functionals of type $\JT{T}{R}{X} = (X \to R) \to T X$. It is assumed throughout that $R$ is a $T$-algebra. We show that $\JT{T}{R}{}$ is also a strong monad, and that it embeds into the continuation monad $\K{R}{X} = (X \to R) \to R$. We use this to derive that the explicitly controlled product of $T$-selection functions is definable from the explicitly controlled product of quantifiers, and hence from Spector's bar recursion. We then prove several properties of this product in the special case when $T$ is the finite power set monad $\power{\cdot}$. These are  used to show that when $T X = \power{X}$ the explicitly controlled product of $T$-selection functions calculates a witness to the Herbrand functional interpretation of the double negation shift.
\end{abstract}

%%%%%%%%%%%%%%%%%%%%%%%%%%%%%%%%%%%%%%%%%%%%%%
%%%%%%%%%%%%%%%%%%%%%%%%%%%%%%%%%%%%%%%%%%%%%%
\section{Introduction}
%%%%%%%%%%%%%%%%%%%%%%%%%%%%%%%%%%%%%%%%%%%%%%
%%%%%%%%%%%%%%%%%%%%%%%%%%%%%%%%%%%%%%%%%%%%%%

G\"odel's \emph{functional} or \emph{Dialectica} interpretation was introduced in \cite{Goedel(58)} as a reduction of first order arithmetic
to the ``finitistic" quantifier-free calculus of primitive recursive functionals (system $\systemT$). Soon after G\"odel's paper appeared in print, Spector
\cite{Spector(62)} showed how G\"odel's interpretation of arithmetic could be extended to analysis by extending
system $\systemT$ with what he called \emph{bar recursion}. By \emph{analysis} we mean
classical arithmetic in all finite types extended with countable choice and dependent choice -- and hence comprehension.

Spector's original work has given rise to several other bar recursive interpretations of analysis, whereby different proof interpretations
other than the Dialetica interpretation have been used. In such cases one was either able to continue using Spector's original form
of bar recursion (e.g. \cite{FE(2010),Kohlenbach(92)}) or some variant of bar recursion was proposed (e.g. \cite{Berardi(98),BO(05)}).

As we have shown in \cite{EO(2009),EO(2011A)}, there are close connections between the different forms of bar recursion and the calculation of optimal strategies in a general class of sequential games. This was achieved by showing that bar recursion turns out to correspond to the iterated product of
quantifiers and selection functions. Spector's original bar recursion can be shown to be equivalent to the iterated product of quantifiers, whereas
the restricted form needed to witness the Dialectica interpretation of $\DNS$ is equivalent to the iterated product of selection functions \cite{EO(2015A)}.

This analogy between computability and games is based on the modelling of players via quantifiers $\K{R}{X} = (X \to R) \to R$. If $X$ is the set of moves available to a player, and $R$ is the set of possible outcomes, then mappings of type $X \to R$ can be seen as describing the context a player lives in. Such contexts (a form of continuation) describe the final outcome for each of the possible choices of the player. Hence, to specify a player is to describe her preferred outcomes for each given game context. Similarly, a selection function $\J{R}{X} = (X \to R) \to X$ also takes a game context as input, but determines the optimal move for any given game context. 

In this paper we consider the iterated product of selection functions parametrised by an arbitrary strong monad $T X$, i.e. $\JT{T}{R}{X} = (X \to R) \to TX$. Using the intuition that an element of a monad $T X$ provides ``information" about concrete elements of $X$, and the correspondence with games, we can view such selection functions $\JT{T}{R}{X}$ as specifying some information about the optimal move for any given game context. 

We study the bar recursion that arises from the iterated product of such $T$-selection functions. Our first step is to show that $\JT{T}{R}{X}$ is also a strong monad. Since any strong monad embeds into the continuation monad, it follows that we have an embedding of $\JT{T}{R}{X}$ into $K X$. We make use of this embedding to show that the iterated
product of $T$-selection functions is in fact primitive recursively definable from the iterated product of quantifiers, and hence from Spector's original bar recursion.

Finally, we consider the particular case when $T X$ is the finite power set monad $\power{X}$. We prove several properties of the iterated product of
selection functions $(X \to R) \to \power{X}$, and show how it provides a witness for the \emph{Herbrand functional interpretation} \cite{BBS(2012)} of double-negation shift $\DNS$
\[
\forall n^\NN \neg \neg A(n) \to \neg \neg \forall n^\NN A(n).
\]
%

%%%%%%%%%%%%%%%%%%%%%%%%%%%%%%%%%%%%%%%%%%%%%%
\subsection{Heyting arithmetic in all finite types, and bar induction}
%%%%%%%%%%%%%%%%%%%%%%%%%%%%%%%%%%%%%%%%%%%%%%

We work in the setting of Heyting arithmetic in all finite types, with full extensionality. This corresponds to the system $\EHAomega$ of \cite{Troelstra(73)}. When carrying out the verification of the Herbrand functional interpretation of $\DNS$ we will make free use of classical logic, in order to simplify the verification of the bar-recursive construction, hence will be working on $\EPAomega$. Although it is well-known that full extensionality is not normally interpreted by the functional interpretations, we are simply assuming full extensionality in the verification of our interpretation of $\DNS$, which is obviously harmless.

The quantifier-free part of the theories $\EHAomega$ and $\EPAomega$ is normally referred to as G\"odel's system $\systemT$. Although in $\systemT$ one normally only assumes the natural numbers $\NN$ as basic types, and function space constructions $X \to Y$ as the only type constructor, we will follow here the same formulation of $\systemT$ as in \cite{BBS(2012)} where one also assumes products $X \times Y$, finite sequences $X^*$, and even finite power sets $\power{X}$. We write $r \preceq s$ to say that the finite sequence $r$ is a prefix of the finite sequence $s$. We assume that each type $X$ contains a `default' value ${\bf 0} \colon X$, so that we can define an canonical extension operation $(\cdot)^+ \colon X^* \to X^\NN$ from finite to infinite sequences, by appending an infinite sequence of default values. For instance, for the natural numbers ${\bf 0}^\NN$ could be the number zero, whereas for $\power{X}$ we can take ${\bf 0}^{\power{X}} = \emptyset$.

On top of $\EHAomega$, in the proofs of Lemmas \ref{bar-full} and \ref{t-spector} will make use of the following form of bar induction: 

\begin{definition}[Bar induction] Let $P(s)$ be a universal formula, and $s \colon X^*$. We say that bar induction holds for $P(s)$ if whenever
\begin{itemize}
	\item $\omega(s^+) < |s|$ implies $P(s)$, and
	\item $\omega(s^+) \geq |s|$ and $\forall x P(s * x)$ implies $P(s)$
\end{itemize}
then $P(\emptyseq)$. 
\end{definition}

This form of bar induction implicitly assumes that the bar condition $\omega(s^+) < |s|$ eventually holds. 
This is indeed the case in all models of Spector's bar recursion \cite{Bezem(85),Scarpellini(71)}. \\[2mm]
{\bf Notation}. In the paper we will use sub-scripts in four different ways, and hope their respective meanings will be clear from context:
\begin{itemize}
	\item In the following section we use sub-scripts to denote the type of a functional. For instance, the identity function of type $X$ will be written as $\id_X$.
	\item If $\alpha \colon X \to Y$ we can view $\alpha$ as a family of elements of $Y$ indexed by $X$, i.e. $\{ \alpha_x \}_{x \colon X}$. When taking this view we might write $\alpha_x$ instead of $\alpha(x)$.
	\item Bar recursive functionals have several parameters, normally $\BR(\omega)(s)(\varepsilon)(q)$. In order to focus on the selection functions $\varepsilon$ and the outcome function $q$ we shall rewrite this as $\BR^\omega_s(\varepsilon)(q)$. This makes sense since $s$ is the `index' of the bar recursion whereas $\omega$ is the stopping condition.
	\item Finally, for $q \colon X^* \to R$ we write $q_s(t)$ for $q(s * t)$ when we wish to `partially evaluate' $q$ on $s$ to produce another function $q_s \colon X^* \to R$.
\end{itemize}

%%%%%%%%%%%%%%%%%%%%%%%%%%%%%%%%%%%%%%%%%%%%%%
\subsection{Strong monads}
%%%%%%%%%%%%%%%%%%%%%%%%%%%%%%%%%%%%%%%%%%%%%%

In this section we recall the basic notions about strong monads needed in this paper. Throughout the paper we work in G\"odel's system $\systemT$. 
Hence, $X, Y$ and $R$ should be viewed as finite types\footnote{It will be clear, however, that what we describe would work more generally in any of the
well-known models of higher-order computability.}. 

\begin{definition}[Strong monad] \label{monad-laws} Let $T$ be a meta-level unary operation on simple types, that we will call a \emph{type operator}. A type operator $T$ is called a \emph{strong monad} if we have a family of closed terms
\eqleft{
\begin{array}{rl}
\eta_X & \colon X \to TX \\[2mm]
(\cdot)^\dagger & \colon (X \to TY) \to (TX \to TY)
\end{array}
}
satisfying (provably in $\systemT$) the laws
\begin{itemize}
	\item[$(i)$] $(\eta_X)^\dagger = \id_{T X}$ \\[-3mm]
	\item[$(ii)$] $g^\dagger \circ \eta_Y = g$ \\[-3mm]
	\item[$(iii)$] $(g^\dagger \circ f)^\dagger = g^\dagger \circ f^\dagger$
\end{itemize}
where $g \colon Y \to T R$ and $f \colon X \to TY$. When several strong monads are involved we shall use the super-script in $\eta^T$ so as to be clear which $\eta$ is being used.  \\[1mm]
Given $f \colon X \to Y$ we define $T f \colon T X \to T Y$ by $T f =
(\eta_Y \circ f)^\dagger$.  The laws for the monad show that this
construction makes $T$ into a functor, that is, $T \id_X = {\rm
  id}_{T X}$, and for $g : Y \to Z$ we have $T(g \circ f) = Tg \circ
Tf$.
\end{definition}

Monads have been extensively studied in category theory \cite{kock:monoidal}, programming language semantics \cite{Moggi(1991)}, and in the functional programming community \cite{Wadler92}. In a monad one would normally have a non-uniform mapping from $f \colon X \to TY$ to $f^\dagger \colon TX \to TY$. The term \emph{strong} here refers to the assumption that we have a uniform map $(\cdot)^\dagger \colon (X \to TY) \to (TX \to TY)$.

\begin{definition}[$T$-algebra] \label{algebra-def} Given a strong monad $T$, a type $R$ is called a \emph{$T$-algebra} if we have a family of maps $(\cdot)^* \colon (X \to R) \to (T X \to R)$ satisfying
\begin{itemize}
	\item[$(i)$] $g^* \circ \eta_Y = g$ \\[-3mm]
	\item[$(ii)$] $(g^* \circ f)^* = g^* \circ f^\dagger$
\end{itemize}
where $g \colon Y \to R$ and $f \colon X \to TY$. 
% For any $T$-algebra $R$ we can define the mapping $\bigcup \colon TR \to R$ as $\bigcup = (\id_R)^*$, and e.g. one can easily show $\bigcup \, \circ \; \eta_R = \id_R$.
% [[ WE SHOULD ACTUALLY TAKE $(\cdot)^* \colon (X \to R) \to (TX \to R)$ AS PRIMITIVE AND HAVE SIMILAR LAWS TO $(\cdot)^\dagger$ ]]
\end{definition}

% Given $f \colon X \to Y$ let $T f \colon T X \to T Y$ be $T f = (\eta_Y \circ f)^\dagger$. As a warm up let us consider the following property, which we will later need in the proof of Lemma \ref{t-spector}.

% \begin{lemma} \label{drop-lemma} If $g \circ f = \id_X$ then $Tg \circ Tf = \id_{T X}$, where $f \colon X \to Y$ and $g \colon Y \to X$.
% \end{lemma}
% %
% {\bf Proof}. Let $(*)$ refer to the assumption $g \circ f = \id_X$. We calculate as follows:
% %
% \eqleft{
% \begin{array}{lcl}
% Tg \circ Tf
% 	& = & (\eta_X \circ g)^\dagger \circ (\eta_Y \circ f)^\dagger \\[2mm]
% 	& \stackrel{\textup{D}\ref{monad-laws}(iii)}{=} & ((\eta_X \circ g)^\dagger \circ \eta_Y \circ f)^\dagger \\[1mm]
% 	& \stackrel{\textup{D}\ref{monad-laws}(ii)}{=} & (\eta_X \circ g \circ f)^\dagger \\[1mm]
% 	& \stackrel{(*)}{=} & (\eta_X \circ \id_X)^\dagger \\[2mm]
% 	& = & (\eta_X)^\dagger \\[1mm]
% 	& \stackrel{\textup{D}\ref{monad-laws}(i)}{=} & \id_{TX}, \\[1mm]
% \end{array}
% }
% where the two unlabelled steps follow by the definition of $T$ and the basic property of the identity functional, respectively.
% $\hfill \Box$

The reason we focus here on \emph{strong} monads is that on such monads we can define a binary product operation as follows:

\begin{lemma} \label{lemma-strong} For any strong monad $T$ we can define a product operation
\[ \otimes \;\colon\; TX \times (X \to TY) \to T(X \times Y) \]
as
\begin{equation} \label{t-product}
a \otimes f = (\lambda x . (\lambda y . \eta_{X \times Y}(x, y))^\dagger(f x))^\dagger(a)
\end{equation}
satisfying, for $q \colon X \times Y \to T R$, 
\[ q^\dagger(a \otimes f) = (\lambda x . (q_x)^\dagger(f x))^\dagger(a), \]
where $q_x = \lambda y . q(x,y)$. When $q \colon X \times Y \to R$ and $R$ is a $T$-algebra it satisfies
\[ q^*(a \otimes f) = (\lambda x . (q_x)^*(f x))^*(a). \]
\end{lemma}
{\bf Proof}. We calculate as follows:
\[
\begin{array}{lcl}
q^\dagger(a \otimes f)
	& \stackrel{(\ref{t-product})}{=} & q^\dagger((\lambda x . (\lambda y . \eta_{X \times Y}(x, y))^\dagger(f x))^\dagger(a)) \\[1mm]
	& \stackrel{(\circ)}{=} & (q^\dagger \circ (\lambda x . (\lambda y . \eta_{X \times Y}(x, y))^\dagger(f x))^\dagger)(a) \\[1mm]
	& \stackrel{\textup{D}\ref{monad-laws}(iii)}{=} & (q^\dagger \circ (\lambda x . (\lambda y . \eta_{X \times Y}(x, y))^\dagger(f x)))^\dagger(a) \\[1mm]
	& \stackrel{(\circ)}{=} & (\lambda x. q^\dagger((\lambda y . \eta_{X \times Y}(x, y))^\dagger(f x)))^\dagger(a) \\[1mm]
	& \stackrel{(\circ)}{=} & (\lambda x. ((q^\dagger \circ (\lambda y . \eta_{X \times Y}(x, y))^\dagger)(f x)))^\dagger(a) \\[1mm]
	& \stackrel{\textup{D}\ref{monad-laws}(iii)}{=} & (\lambda x. ((q^\dagger \circ (\lambda y . \eta_{X \times Y}(x, y)))^\dagger(f x)))^\dagger(a) \\[1mm]
	& \stackrel{(\circ)}{=} & (\lambda x. ((    \lambda y . q^\dagger (\eta_{X \times Y}(x, y))     )^\dagger(f x)))^\dagger(a) \\[1mm]
	& \stackrel{\textup{D}\ref{monad-laws}(ii)}{=} & (\lambda x . (q_x)^\dagger(f x))^\dagger(a).
\end{array}
\]
In the case $q \colon X \times Y \to R$ and $R$ is a $T$-algebra we use properties $(i)$ and $(ii)$ of Definition \ref{algebra-def} instead.
$\hfill \Box$

%%%%%%%%%%%%%%%%%%%%%%%%%%%%%%%%%%%%%%%%%%%%%%
%%%%%%%%%%%%%%%%%%%%%%%%%%%%%%%%%%%%%%%%%%%%%%
\section{$T$-Selection Functions}
%%%%%%%%%%%%%%%%%%%%%%%%%%%%%%%%%%%%%%%%%%%%%%
%%%%%%%%%%%%%%%%%%%%%%%%%%%%%%%%%%%%%%%%%%%%%%

In the following two sections we assume that $T$ is a strong monad, and that $R$ is a $T$-algebra.

\begin{definition}[$T$-selection functions] Let $\JT{T}{R}{X} = (X \to R) \to T X$, where $R$ is a $T$-algebra. The elements of the type $\JT{T}{R}{X}$ will be called \emph{$T$-selection functions}. 
\end{definition}

Under the assumptions that $T$ is a strong monad and $R$ a $T$-algebra, it follows that $\JT{T}{R}{}$ is also a strong monad.

\begin{lemma} \label{j-t-monad} $\JT{T}{R}{}$ is a strong monad with operations:
\eqleft{
\begin{array}{cl}
(i) & \eta^{\JT{T}{R}{}}_X(x) = \lambda p . \eta^T_X(x) \\[2mm]
(ii) & \delta^\dagger(\varepsilon) = \lambda p . (b^\delta_p)^\dagger (a^{\varepsilon,\delta}_p), \mbox{where $\delta \colon X \to \JT{T}{R}{Y}$ and $\delta^\dagger \colon \JT{T}{R}{X} \to \JT{T}{R}{Y}$}
\end{array}
}
where $b_p^\delta(x) \stackrel{T Y}{=} \delta(x)(p)$ and $a^{\varepsilon,\delta}_p \stackrel{T X}{=} \varepsilon(p^* \circ b_p^\delta)$. 
\end{lemma}
{\bf Proof}. It is easy to check conditions ($i$) and ($ii$). Define $\Delta_x(p) = (b^\delta_p)^\dagger (a^{\varepsilon_x,\delta}_p)$ and $\Gamma_\nu(p) = (b^\varepsilon_p)^\dagger(a^{\nu, \varepsilon}_p)$. We outline property ($iii$):
\[
\begin{array}{lcl}
(\delta^\dagger \circ \varepsilon)^\dagger 
	& = & (\lambda x . \delta^\dagger(\varepsilon_x))^\dagger \\[0mm]
	& \stackrel{(ii)}{=} & (\lambda x . \lambda p . (b^\delta_p)^\dagger (a^{\varepsilon_x,\delta}_p))^\dagger \\[0mm]
	& \stackrel{\Delta\;\textup{def.}}{=} & (\lambda x . \Delta_x)^\dagger \\[0mm]
	& \stackrel{(ii)}{=} & \lambda \nu . \lambda q . (b_q^{\Delta})^\dagger(a_q^{\nu, \Delta}) \\[1mm]
	& \stackrel{b\;\textup{def.}}{=} & \lambda \nu . \lambda q . (\lambda x . \Delta_x(q))^\dagger(a_q^{\nu, \Delta}) \\[0mm]
	& \stackrel{\Delta\;\textup{def.}}{=} & \lambda \nu . \lambda q . ((b^\delta_q)^\dagger \circ (\lambda x . a^{\varepsilon_x,\delta}_q))^\dagger(a_q^{\nu, \Delta}) \\[0mm]
	& \stackrel{\textup{D}\ref{monad-laws}(iii)}{=} & \lambda \nu . \lambda q . ((b^\delta_q)^\dagger \circ (\lambda x . a^{\varepsilon_x,\delta}_q)^\dagger)(a_q^{\nu, \Delta}) \\[1mm]
	& = & \lambda \nu . \lambda q . (b^\delta_q)^\dagger ((\lambda x . a^{\varepsilon_x,\delta}_q)^\dagger(a_q^{\nu, \Delta})) \\[0mm]
	& \stackrel{(*)}{=} & \lambda \nu . \lambda q . (b_q^\delta)^\dagger (a^{\Gamma_\nu, \delta}_q) \\[0mm]
	& \stackrel{(ii)}{=} & \lambda \nu . \delta^\dagger(\Gamma_\nu) \\[0mm]
	& \stackrel{\Gamma\;\textup{def.}}{=} & \lambda \nu . \delta^\dagger(\lambda p . (b^\varepsilon_p)^\dagger(a^{\nu, \varepsilon}_p)) \\[0mm]
	& \stackrel{(ii)}{=} & \lambda \nu . \delta^\dagger(\varepsilon^\dagger(\nu)) \\[1mm]
	& = & \delta^\dagger \circ \varepsilon^\dagger.
\end{array}
\]
It remains to show that $(*) \; a^{\Gamma_\nu, \delta}_q = (\lambda x . a^{\varepsilon_x,\delta}_q)^\dagger(a_q^{\nu, \Delta})$. This can be shown as
\[
\begin{array}{lcl}
a^{\Gamma_\nu, \delta}_q  
	& \stackrel{a\;\textup{def.}}{=} & \Gamma_\nu(q^* \circ b_q^\delta) \\[0mm]
	& \stackrel{\Gamma\;\textup{def.}}{=} & (\lambda x . b^\varepsilon_{q^* \circ b_q^\delta}(x))^\dagger(a^{\nu, \varepsilon}_{q^* \circ b_q^\delta}) \\[0mm]
	& \stackrel{b\;\textup{def.}}{=} & (\lambda x . \varepsilon_x(q^* \circ b_q^\delta))^\dagger(a^{\nu,\varepsilon}_{q^* \circ b_q^\delta}) \\[0mm]
	& \stackrel{(**)}{=} & (\lambda x . \varepsilon_x (q^* \circ b_q^\delta))^\dagger(a_q^{\nu, \Delta}) \\[0mm]
	& \stackrel{a\;\textup{def.}}{=} & (\lambda x . a^{\varepsilon_x,\delta}_q)^\dagger(a_q^{\nu, \Delta}),
\end{array}
\]
where, finally, $(**) \; a^{\nu,\varepsilon}_{q^* \circ b_q^\delta} = a_q^{\nu, \Delta}$ is shown as
\[
\begin{array}{lcl}
a^{\nu,\varepsilon}_{q^* \circ b_q^\delta} 
	& \stackrel{a\;\textup{def.}}{=} & \nu((q^* \circ b_q^\delta)^* \circ b_{q^* \circ b_q^\delta}^\varepsilon) \\[0mm]
	& \stackrel{\textup{D}\ref{algebra-def}(ii)}{=} & \nu(q^* \circ (b_q^\delta)^\dagger \circ b_{q^* \circ b_q^\delta}^\varepsilon) \\[0mm]
	& = & \nu(\lambda x . q^*((b_q^\delta)^\dagger(b_{q^* \circ b_q^\delta}^\varepsilon(x)))) \\[0mm]
	& \stackrel{b\;\textup{def.}}{=} & \nu(\lambda x . q^*((b^\delta_q)^\dagger (\varepsilon_x(q^* \circ b_q^\delta)))) \\[0mm]
	& \stackrel{a\;\textup{def.}}{=} & \nu(\lambda x . q^*((b^\delta_q)^\dagger (a^{\varepsilon_x,\delta}_q))) \\[0mm]
	& \stackrel{\Delta\;\textup{def.}}{=} & \nu(\lambda x . q^*(\Delta_x(q))) \\[0mm]
	& \stackrel{b\;\textup{def.}}{=} & \nu(\lambda x . q^*(b_q^\Delta(x))) \\[0mm]
	& \stackrel{a\;\textup{def.}}{=} & a_q^{\nu, \Delta}.
\end{array}
\]
$\hfill \Box$

It follows that the product operation of the monad $\JT{T}{R}{}$ can be explicitly described in terms of the product operation on $T$ as:
\begin{equation} \label{j-t-monad-eq}
(\varepsilon \otimes^{\JT{T}{R}{}} \delta)(q) = a \otimes^T f 
\end{equation}
where $q \colon X \times Y \to R$, $\varepsilon \colon (X \to R) \to T X$ and $\delta \colon X \to (Y \to R) \to T Y$, and
\eqleft{
\begin{array}{lcl}
	f(x) & \stackrel{TY}{=} & \delta_x(q_x) \\[1mm]
	a & \stackrel{TX}{=} & \varepsilon(\lambda x^X. (q_x)^*(f x)).
\end{array}
}

Note that $\otimes$ on the right side of (\ref{j-t-monad-eq}) denotes the product on the strong monad $T$ whereas $\otimes$ on the left denotes the product of the strong monad $\JT{T}{R}{}$. We will in general use the same notation $\otimes$ for the product of any strong monad, as it will hopefully be clear from the context which monad we are referring to.

\begin{definition}[from $\JT{T}{R}{}$ to $\K{R}{}$] \label{bar-def} Let $\K{R}{X} = (X \to R) \to R$. Given a $T$-selection function $\varepsilon \colon \JT{T}{R}{X}$ we can construct a quantifier $\overline{\varepsilon} \colon \K{R}{X}$ as
\[ \overline{\varepsilon}(p^{X \to R}) \;\stackrel{R}{=}\; p^*(\varepsilon p). \]
\end{definition}

It can be shown that the construction $\varepsilon \mapsto \overline{\varepsilon}$ is actually a monad morphism, from which the next lemma follows. Nevertheless, we shall prove the lemma directly. A particular instance of this lemma, when $T$ is the identity monad, was first proven in \cite{EO(2009)}. It is important here that $R$ is a $T$-algebra.

\begin{lemma} \label{bar-binary} Given $\varepsilon \colon \JT{T}{R}{X}$ and $\delta \colon X \to \JT{T}{R}{Y}$ then
\[ \overline{(\varepsilon \otimes^{\JT{T}{R}{}} \delta)} =  \overline{\varepsilon} \otimes^{\K{R}{}} (\lambda x . \overline{\delta_x}). \]
\end{lemma}
{\bf Proof}. Define $f(x) = \delta_x(q_x)$ and $p(x) = (q_x)^*(f x)$ and $a = \varepsilon(p)$. We calculate as follows:
\eqleft{
\begin{array}{lcl}
\overline{(\varepsilon \otimes^{\JT{T}{R}{}} \delta)}(q)
	& \stackrel{\textup{D}\ref{bar-def}}{=} & q^*((\varepsilon \otimes^{\JT{T}{R}{}} \delta)(q)) \\[0mm]
	& \stackrel{(\ref{j-t-monad-eq})}{=} & q^*(a \otimes^T f) \\[1mm]
	& \stackrel{\textup{L}\ref{lemma-strong}}{=} & (\lambda x . (q_x)^*(f x))^*(a) \\[1mm]
	& \stackrel{\textup{Def($a$)}}{=} & (\lambda x . (q_x)^*(f x))^*(\varepsilon(p)) \\[1mm]
	& \stackrel{\textup{Def($p$)}}{=} & p^*(\varepsilon(p)) \\[1mm]
	& \stackrel{\textup{D}\ref{bar-def}}{=} & \overline{\varepsilon}(p) \\[1mm]
	& \stackrel{\textup{Def($p, f$)}}{=} & \overline{\varepsilon}(\lambda x . (q_x)^*(\delta_x(q_x))) \\[1mm]
	& \stackrel{\textup{D}\ref{bar-def}}{=} & \overline{\varepsilon}(\lambda x . \overline{\delta_x}(q_x)) \\[2mm]
	& = & (\overline{\varepsilon} \otimes^{\K{R}{}} (\lambda x . \overline{\delta_x}))(q).
\end{array}
}
The last equality in the chain above uses the definition of the product $\otimes$ for the strong monad $\K{R}{X}$. 
$\hfill \Box$

%%%%%%%%%%%%%%%%%%%%%%%%%%%%%%%%%%%%%%%%%%%%%%
%%%%%%%%%%%%%%%%%%%%%%%%%%%%%%%%%%%%%%%%%%%%%%
\section{Iterated Products and Bar Recursion}
%%%%%%%%%%%%%%%%%%%%%%%%%%%%%%%%%%%%%%%%%%%%%%
%%%%%%%%%%%%%%%%%%%%%%%%%%%%%%%%%%%%%%%%%%%%%%

Given any strong monad $M$ we can iterate its product operation $M X \times (X \to M Y) \to M(X \times Y)$ so as to obtain an operation\footnote{A simpler instance of this operation without dependent types, namely $(M X)^\NN \to M(X^\NN)$, is actually a built-in function in standard implementations of the Haskell programming language called \sf{sequence :: Monad m =$>$ [m a] -$>$ m [a]}.} on infinite sequences $(X^* \to M(X))^\NN \to M(X^\NN)$. Although this will not be a total operation in general, it is surprising that, as shown in \cite{EO(2009)}, it defines a total operation when $M$ is the selection monad $M X = \J{R}{X}$ and $R$ is a discrete type. 

It is also possible to iterate the binary product of $M$ in a controlled way, by using an explicit termination function $\omega \colon X^\NN \to \NN$ as
\[
\IEP{M}^\omega_s(\alpha) = 
\left\{
\begin{array}{ll}
\eta^M(\emptyseq) & {\rm if} \; \omega(s^+) < |s| \\[2mm]
\alpha_s \otimes^M (\lambda x . \IEP{M}^\omega_{s * x}(\alpha) & {\rm otherwise}
\end{array}
\right.
\]
where $\IEP{M}^\omega_s$ is of type $(\NN \to M(X) \to M(X^*)$. We use the acronym $\IEP{M}$ for the ``explicitly controlled iterated product of the strong monad $M$''.

The explicitly controlled product of selection functions $\EPS$ or quantifiers $\EPQ$ (cf. \cite{EO(2015A)}) are particular cases when $M X = \J{R}{X}$ and $M X = \K{R}{X}$, this time for an arbitrary $R$, i.e. $\EPS = \IEP{\J{R}{}}$ and $\EPQ = \IEP{\K{R}{}}$. In turn, these are primitively recursively equivalent to restricted Spector bar recursion and the general Spector bar recursion, respectively \cite{EO(2015A)}.

In this section we consider another instance where $M X = \JT{T}{R}{X}$, with $T$ being a strong monad, i.e. $\IEP{\JT{T}{R}{}}$ which we shall call $\TEPS$.

\begin{definition}[Iterated $\JT{T}{R}{}$ product] \label{fin-set-br} Let $\varepsilon_s \colon \JT{T}{R}{X}_{|s|}$ and $s \colon X^*$ and $\omega \colon X^\NN \to \NN$. 
We define $\TEPS^\omega_s(\varepsilon) \colon \JT{T}{R}{X}^*$ as $\TEPS^\omega_s = \IEP{\JT{T}{R}{}}^\omega_s$.
%
%\[
%\TEPS^\omega_s(\varepsilon) = 
%\left\{
%\begin{array}{ll}
%\eta^{\JT{T}{R}{}}(\emptyseq) & {\rm if} \; \omega(s^+) < |s| \\[2mm]
%\varepsilon_s \otimes^{\JT{T}{R}{}} \lambda x . \TEPS^\omega_{s * x}(\varepsilon) & {\rm otherwise}.
%\end{array}
%\right.
%\]
%
\end{definition}

Unfolding the definition of the binary product, as in Lemma \ref{j-t-monad}, and noticing that $\eta^{\JT{T}{R}{}}(\emptyseq) = \lambda q . \eta^T(\emptyseq)$, the equation above can be also written as
\begin{equation} \label{monad-br}
\TEPS^\omega_s(\varepsilon)(q) = 
\left\{
\begin{array}{ll}
\eta^T(\emptyseq) & {\rm if} \; \omega(s^+) < |s| \\[2mm]
a \otimes^T f & {\rm otherwise}
\end{array}
\right.
\end{equation}
where $a = \varepsilon_s(\lambda x . (q_x)^*(f x))$ and $f(x) = \TEPS^\omega_{s * x}(\varepsilon)(q_x)$. 

Recall that $\EPQ$ is the explicitly controlled iterated product of quantifiers, i.e. $\EPQ = \IEP{\K{R}{}}$. $\EPQ$ satisfies the equation
\[
\EPQ^\omega_s(\phi) = 
\left\{
\begin{array}{ll}
\lambda q . q(\emptyseq) & {\rm if} \; \omega(s^+) < |s| \\[2mm]
\phi_s \otimes^{\K{R}{}} \lambda x . \EPQ^\omega_{s * x}(\phi) & {\rm otherwise}.
\end{array}
\right.
\]
Again, the definition of the binary product of quantifiers can unfolded, leading to the equivalent equation
\begin{equation} \label{epq-eq}
\EPQ^\omega_s(\phi)(q) = 
\left\{
\begin{array}{ll}
q(\emptyseq) & {\rm if} \; \omega(s^+) < |s| \\[2mm]
\phi_s(\lambda x. \EPQ^\omega_{s * x}(\phi)(q_x)) & {\rm otherwise}
\end{array}
\right.
\end{equation}
As show in \cite{EO(2010A)}, $\EPQ$ is equivalent over system $\systemT$ to Spector's bar recursion. The following lemma follows by a simple iteration of  Lemma \ref{bar-binary}. 

\begin{lemma} \label{bar-full} $\overline{\TEPS^\omega_{\emptyseq}(\varepsilon)} = \EPQ^\omega_{\emptyseq}(\overline{\varepsilon})$.
\end{lemma}
{\bf Proof}. The proof goes by bar induction on $s$ with the bar $\omega(s^+) < |s|$.  In case we have reached the bar, i.e. $\omega(s^+) < |s|$, we have
\eqleft{
\begin{array}{lcl}
\EPQ^\omega_s(\overline{\varepsilon})(q) 
	& = & q(\emptyseq) \\[1mm]
	& \stackrel{\textup{D}\ref{algebra-def}(i)}{=} & q^*(\eta^T(\emptyseq)) \\[1mm]
	& \stackrel{\textup{D}\ref{fin-set-br}}{=} & q^*(\TEPS^\omega_s(\varepsilon)(q)) \\[1mm]
	& \stackrel{\textup{D}\ref{bar-def}}{=} & \overline{\TEPS^\omega_s(\varepsilon)}.
\end{array}
}
By the bar inductive assumption we have that $\overline{\TEPS^\omega_{s * x}(\varepsilon)} = \EPQ^\omega_{s * x}(\overline{\varepsilon})$, for all $x$, and hence
\eqleft{
\begin{array}{lcl}
\EPQ^\omega_s(\overline{\varepsilon})(q) 
	& = & (\overline{\varepsilon} \otimes^{\K{R}{}} (\lambda x . \EPQ^\omega_{s * x}(\overline{\varepsilon})))(q) \\[2mm]
	& \stackrel{\textup{(IH)}}{=} & (\overline{\varepsilon} \otimes^{\K{R}{}} (\lambda x . \overline{\TEPS^\omega_{s * x}(\varepsilon)}))(q) \\[2mm]
	& \stackrel{\textup{L}\ref{bar-binary}}{=} & (\overline{\varepsilon \otimes^{\JT{T}{R}{}} (\lambda x . \TEPS^\omega_{s * x}(\varepsilon))})(q) \\[2mm]
	& = & \overline{\TEPS^\omega_s(\varepsilon)}(q), \\[2mm]
\end{array}
}
since we can assume $\omega(s^+) \geq |s|$.
$\hfill \Box$

It is well know that the product of selection functions of type $(X, R)$ can be simulated by a product where $R$ is restricted to $R = X^\NN$ and $q \colon X^\NN \to R$ is the identity function. In fact, one can think of Spector's restricted form of bar recursion \cite{Spector(62)} as the iterated product of these restricted selection functions. In terms of games, it corresponds to taking the outcome of the game to be the sequence of moves played. The actual outcome of the game can be reconstructed from this sequence via the outcome function. The next lemma shows that this simulation of an arbitrary outcome type $R$ by taking the outcome to be the actual sequence of moves also works in this monadic setting.

\begin{lemma} \label{t-spector} $\TEPS$ of type $(X, R)$ is definable from $\TEPS$ of type $(X, T X^\NN)$.
\end{lemma}
{\bf Proof}. Let ${\sf add}_s \colon X^\NN \to X^\NN$ and ${\sf drop}_n \colon X^\NN \to X^\NN$ be the functions that append the finite sequence $s$ to the beginning of an infinite list, and the function that drops $n$ elements from an infinite list, respectively. Clearly, ${\sf drop}_{|s|} \circ {\sf add}_s$ is the identity, and hence, by functoriality, $T({\sf drop}_{|s|}) \circ T({\sf add}_s)$ is the identity on $T(X^\NN)$. Given $q \colon X^\NN \to R$ and $\varepsilon_s \colon \JT{T}{R}{X}$ we define $\varepsilon^q_s \colon \JT{T}{T X^\NN}{X}$ as
\[ \varepsilon^q_s(p^{X \to T X^\NN}) \stackrel{T X}{=} \varepsilon_s(\lambda x . ((q_{s * x})^* \circ T({\sf drop}_{|s*x|}))(p x)). \]
Note that $T X^\NN$ is also a $T$-algebra with the map
\[ (\cdot)^* \colon (Y \to T X^\NN) \to (T Y \to T X^\NN) \]
being simply the $(\cdot)^\dagger$ of the monad $T$. We claim that \[ \TEPS^\omega_{\emptyseq}(\varepsilon)(q) = \TEPS^\omega_{\emptyseq}(\varepsilon^q)(\eta^T). \] Define
\eqleft{P(s) \equiv \TEPS^\omega_s(\varepsilon)(q_s) = \TEPS^\omega_s(\varepsilon^q)(\eta^T \circ {\sf add}_s)}
and let us show $P(\emptyseq)$ by bar induction. Recall that $T({\sf add}_s) = \eta^T \circ {\sf add}_s$ by definition. In the base case, assuming $\omega(s^+) < |s|$, we have
\eqleft{\TEPS^\omega_s(\varepsilon)(q_s) = \eta^T(\emptyseq) =  \TEPS^\omega_s(\varepsilon^q)(\eta^T \circ {\sf add}_s).}
For the bar inductive step we assume $P(s * x)$ holds for all $x$ and must prove $P(s)$. We can also assume that $\omega(s^+) \geq |s|$. Let 
\eqleft{
\begin{array}{rcl}
	f(x) & = & \TEPS^\omega_{s * x}(\varepsilon)(q_{s * x}) \\[2mm]
	a & = & \varepsilon_s(\lambda x . (q_{s * x})^*(f x)) \\[2mm]
	\tilde{f}(x) & = & \TEPS^\omega_{s * x}(\varepsilon^q)(\eta^T \circ {\sf add}_{s * x}) \\[2mm]
	\tilde{a} & = & \varepsilon_s^q(\lambda x . T({\sf add}_{s * x})(\tilde{f} x)).
\end{array}
}
By the bar inductive hypothesis we have $f = \tilde{f}$ and hence
\eqleft{
\begin{array}{lcl}
\tilde{a}
	& = & \varepsilon_s^q(\lambda x . T({\sf add}_{s * x})(\tilde{f} x)) \\[1mm]
	& \stackrel{\textup{(IH)}}{=} & \varepsilon_s^q(\lambda x . T({\sf add}_{s * x})(f x)) \\[1mm]
	& \stackrel{(\varepsilon^q\,\textup{def})}{=} & \varepsilon_s(\lambda x . ((q_{s * x})^* \circ T({\sf drop}_{|s * x|}))(T({\sf add}_{s * x})(f x)))) \\[2mm]
	& = & \varepsilon_s(\lambda x . (q_{s*x})^*(fx)) \\[2mm]
	& = & a.
\end{array}
}
Therefore
\eqleft{
\begin{array}{lcl}
\TEPS^\omega_s(\varepsilon)(q_s)
	& = & a \otimes^T f \\[2mm]
	& = & \tilde{a} \otimes^T \tilde{f} \\[2mm]
	& = & \TEPS^\omega_s(\varepsilon^q)(\eta^T \circ {\sf add}_{s}).
\end{array}
}
In the last step we have used that $T({\sf add}_s)$ is defined as $\eta^T \circ {\sf add}_s$.
$\hfill \Box$

The main result in this section is that Spector's original bar recursion already defines the explicitly controlled product  of $T$-selection functions $\TEPS$. Spector proves this in \cite{Spector(62)} for the case when $T$ is the identity monad. The following theorem shows that this in fact holds for any strong monad $T$. 

\begin{theorem} \label{thm-teps-from-epq} $\TEPS$ is definable from $\EPQ$.
\end{theorem}
{\bf Proof}. We claim that $\TEPS^\omega_{\emptyseq}(\varepsilon)(q)$ can be defined as $\EPQ^\omega_{\emptyseq}(\overline{\varepsilon^q})(\eta)$, where $\varepsilon^q$ is as in the proof of the previous lemma. Indeed we have:
\[
\begin{array}{lcl}
\EPQ^\omega_{\emptyseq}(\overline{\varepsilon^q})(\eta)
	& \stackrel{\textup{L}\ref{bar-full}}{=} & \overline{\TEPS^\omega_{\emptyseq}(\varepsilon^q)}(\eta) \\[1mm]
	& \stackrel{\textup{D}\ref{bar-def}}{=} & \eta^*(\TEPS^\omega_{\emptyseq}(\varepsilon^q)(\eta)) \\[1mm]
	& \stackrel{\textup{L}\ref{t-spector}}{=} & \eta^*(\TEPS^\omega_{\emptyseq}(\varepsilon)(q)) \\[2mm]
	& = & \eta^\dagger(\TEPS^\omega_{\emptyseq}(\varepsilon)(q)) \\[1mm]
	& \stackrel{\textup{D}\ref{monad-laws}(i)}{=} & \TEPS^\omega_{\emptyseq}(\varepsilon)(q).
\end{array}
\]
We used that the map $(\cdot)^*$ for the algebra $T X^\NN$ is just the $(\cdot)^\dagger$ map for the monad $T$, as discussed in the proof of Lemma \ref{t-spector}.
$\hfill \Box$

%%%%%%%%%%%%%%%%%%%%%%%%%%%%%%%%%%%%%%%%%%%%%%
%%%%%%%%%%%%%%%%%%%%%%%%%%%%%%%%%%%%%%%%%%%%%%
\section{Finite Power Sets}
%%%%%%%%%%%%%%%%%%%%%%%%%%%%%%%%%%%%%%%%%%%%%%
%%%%%%%%%%%%%%%%%%%%%%%%%%%%%%%%%%%%%%%%%%%%%%
\label{sec-lemmas}

For the rest of the paper we will make essential use of the definitional extension of G\"odel's system $\systemT$ with the finite power-set type $\power{X}$. To simplify the exposition, let us also abbreviate $\power{X \to Y}$ as $X \Rightarrow Y$, i.e. the type of finite sets of functions from $X$ to $Y$. We can think of the elements $f \colon X \Rightarrow \power{Y}$ as functions by defining the following \emph{set-application} 
\[ \monApp{f}{x^X} \stackrel{\power{Y}}{=} \bigcup_{g \in f} g x. \]
Hence, if $f \colon X \Rightarrow \power{Y}$ then $\monApp{f}{\cdot} \colon X \to \power{Y}$. In particular, if $f \colon (X \Rightarrow (Y \Rightarrow \power{Z}))$ then $\monApp{\monApp{f}{x}}{y}$ stands for
\[ \bigcup_{g \in f} \bigcup_{h \in g x} h y \]
and we will be abbreviated that as $\monAppD{f}{x}{y}$.

\begin{lemma} The finite power set type operator $\power{\cdot}$ is a strong monad with operations
\begin{itemize}
	\item $\eta(x) = \{ x \}$ \\[-2mm]
	\item $f^\dagger(S) = \bigcup \{ f(x) \; \colon \; x \in S \}$, for $f \colon X \to \power{Y}$.
\end{itemize}
Moreover, its binary product
\[ \otimes \colon \power{X} \times (X \to \power{Y}) \to \power{X \times Y} \]
can be explicitly described as
\[ S \otimes f = \{ \langle a, b \rangle \; \colon \; a \in S \wedge b \in f(a) \}. \]
\end{lemma}

For the rest of the paper we shall assume that $R = \power{R'}$, for some $R'$, so that $R$ is an algebra for $\power{\cdot}$ with $(\cdot)^* = (\cdot)^\dagger$. We will also use $\bigcup \colon \power{R} \to R$, the usual union operation which satisfies $S_i \subseteq \bigcup \{ S_i \; \colon \; i \in I \}$  (we use this in Lemma \ref{lemma-main-eq}). 

\begin{definition}[Herbrand bar recursion] Let us write $\hBR$ for the instance of $\TEPS$ where $T = \power{\cdot}$, i.e
\[
\hBR^\omega_s(\varepsilon)(q) = 
\left\{
\begin{array}{ll}
\{\emptyseq\} & {\rm if} \; \omega(s^+) < |s| \\[2mm]
\{ a * r \; \colon \; a \in \chi \wedge r \in \hBR_{s * a}(\omega)(\varepsilon)(q_a) \} & {\rm otherwise}
\end{array}
\right.
\]
where $\chi = \varepsilon_s(\lambda x . \bigcup \{ q_x(r) \; \colon \; r \in \hBR_{s * x}(\omega)(\varepsilon)(q_x) \})$.
\end{definition}

By Theorem \ref{thm-teps-from-epq} $\hBR$ is $T$-definable from Spector's general form of bar recursion \cite{Oliva(2012A)}. We now prove four lemmas about $\hBR$, to be used in the interpretation of $\DNS$ in the following section. For this section we will assume that $\varepsilon$ and $\omega$ are fixed functionals and hence, for the sake of readability, we shall omit these as parameters in $\hBR_s^\omega(\varepsilon)(q)$.

\begin{lemma} \label{lemma-basic} Let $t = \hBR_{\emptyseq}(q)$ and $s \in t$. For all $i \leq |s|$ we have
\eqleft{s \in \{ \langle s_0, \ldots, s_{i-1} \rangle * r \; \colon \; r \in \hBR_{\langle s_0, \ldots, s_{i-1} \rangle}(q_{\langle s_0, \ldots, s_{i-1} \rangle}) \}.}
The types are $t \colon \power{X^*}$ and $s \colon X^*$.
\end{lemma}
{\bf Proof}. By induction on $i$. If $i = 0$ then $\langle s_0, \ldots, s_{i-1} \rangle$ is the empty sequence and the result follows by the assumption that $s \in t$. For the induction step assume that $i < |s|$ and that
\eqleft{s \in \{ \langle s_0, \ldots, s_{i-1} \rangle * r \; \colon \; r \in \hBR_{\langle s_0, \ldots, s_{i-1} \rangle}(q_{\langle s_0, \ldots, s_{i-1} \rangle}) \}.}
Since $i < |s|$ there must exist some $r \in \hBR_{\langle s_0, \ldots, s_{i-1} \rangle}(q_{\langle s_0, \ldots, s_{i-1} \rangle})$ of the form $s_i * r'$ so that
\begin{itemize}
	\item[($i$)] $s = \langle s_0, \ldots, s_{i-1}, s_i \rangle * r'$, and \\[-2mm]
	\item[($ii$)] $s_i * r' \in \hBR_{\langle s_0, \ldots, s_{i-1} \rangle}(q_{\langle s_0, \ldots, s_{i-1} \rangle})$.
\end{itemize}
In particular, we cannot have $\hBR_{\langle s_0, \ldots, s_{i-1} \rangle}(q_{\langle s_0, \ldots, s_{i-1} \rangle}) = \{ \emptyseq \}$, so it must be the case that $(*) \; \omega(\langle s_0, \ldots, s_{i-1} \rangle^+) \geq |\langle s_0, \ldots, s_{i-1} \rangle |$. Hence
\[ \hBR_{\langle s_0, \ldots, s_{i-1} \rangle}(q_{\langle s_0, \ldots, s_{i-1} \rangle}) = \{ a * r \; \colon \; a \in \chi \wedge r \in \hBR_{\langle s_0, \ldots, s_{i-1}, a \rangle}(q_{\langle s_0, \ldots, s_{i-1},a \rangle}) \} \]
where
\[ \chi = \varepsilon_{\langle s_0, \ldots, s_{i-1} \rangle}( \lambda y^{X} . \bigcup \{ q_{\langle s_0, \ldots, s_{i-1}, y \rangle}(r) \; \colon \; r \in \hBR_{\langle s_0, \ldots, s_{i-1}, y \rangle}(q_{\langle s_0, \ldots, s_{i-1}, y \rangle}) \} ). \]
From ($ii$) it follows that $s_i \in \chi$ and
\begin{itemize}
	\item[($iii$)] $r' \in \hBR_{\langle s_0, \ldots, s_{i-1}, s_i \rangle}(q_{\langle s_0, \ldots, s_{i-1}, s_i \rangle})$.
\end{itemize}
Finally, from ($i$) and ($iii$) we have
\eqleft{s \in \{ \langle s_0, \ldots, s_{i-1}, s_i \rangle * r' \; \colon \; r' \in \hBR_{\langle s_0, \ldots, s_{i-1}, s_i \rangle}(q_{\langle s_0, \ldots, s_{i-1}, s_i \rangle}) \}}
which concludes the proof.
$\hfill \Box$

%%%%%%%%%%%%%%%%%%%%%%%%%%%%%%%%%%%%%%%%%%%%%%%%%%%%%%%%%%%%%
%  This is the lemma corresponding to the main lemma on Spector's bar recursion (which solves Spector's equstions)      %
%%%%%%%%%%%%%%%%%%%%%%%%%%%%%%%%%%%%%%%%%%%%%%%%%%%%%%%%%%%%%

For the following three lemmas let $t = \hBR_{\emptyseq}(q)$, and assume $a_0, \ldots, a_n$ is a finite sequence satisfying, for all $i \leq n$,
\eqleft{
a_i \in \varepsilon_{\langle a_0, \ldots, a_{i-1} \rangle}(p_i)
}
where $p_i$ is defined as
\eqleft{
p_i(y) = \bigcup \{ q_{\langle a_0, \ldots, a_{i-1}, y \rangle}(r) \; \colon \; r \in \hBR_{\langle a_0, \ldots, a_{i-1}, y \rangle}(q_{\langle a_0, \ldots, a_{i-1}, y \rangle}) \}.
}

\begin{lemma} \label{lemma-pre-eq-basic} If $\omega(\langle a_0, \ldots, a_{i-1} \rangle^+) \geq i$, for all $i \leq n$, then
\eqleft{\langle a_0, \ldots, a_{n-1} \rangle * x * r \in t,}
for all $x \in \varepsilon_{\langle a_0, \ldots, a_{n-1} \rangle}(p_n)$ and $r \in \hBR_{\langle a_0, \ldots, a_{n-1}, x \rangle}(q_{\langle a_0, \ldots, a_{n-1}, x \rangle})$.
\end{lemma}
{\bf Proof}. We prove the lemma by induction on $n$. \\[1mm]
For $n = 0$ the assumption of the lemma always holds, while the conclusion follows by the definition of $\hBR$ 
\eqleft{t = \hBR_{\emptyseq}(q) = \{ a * r \; \colon \; a \in \varepsilon_{\langle \, \rangle}(p_0) \wedge r \in \hBR_{a}(q_a) \}}
since $\omega(\emptyseq^+) \geq 0$. For the induction step, assume that $\omega(\langle a_0, \ldots, a_{i-1} \rangle^+) \geq i$, for all $i \leq n+1$. In particular this holds for $i \leq n$. Hence, by induction hypothesis we have
\begin{itemize}
	\item[] $\langle a_0, \ldots, a_{n-1} \rangle * x * r \in t$, \\[-3mm]
	\item[] \quad for all $x \in \varepsilon_{\langle a_0, \ldots, a_{n-1} \rangle}(p_n)$ and $r \in \hBR_{\langle a_0, \ldots, a_{n-1}, x \rangle}(q_{\langle a_0, \ldots, a_{n-1}, x \rangle})$
\end{itemize}
and, since $a_n \in \varepsilon_{\langle a_0, \ldots, a_{n-1} \rangle}(p_n)$,
\begin{itemize}
	\item[($i$)] $\langle a_0, \ldots, a_n \rangle * r \in t$, for all $r \in \hBR_{\langle a_0, \ldots, a_n \rangle}(q_{\langle a_0, \ldots, a_n \rangle})$.
\end{itemize}
Now fix a $y \in \varepsilon_{\langle a_0, \ldots, a_n \rangle}(p_{n+1})$ and an $r' \in \hBR_{\langle a_0, \ldots, a_n, y \rangle}(q_{\langle a_0, \ldots, a_n, y \rangle})$. In order to show that $\langle a_0, \ldots, a_n \rangle * y * r' \in t$, by ($i$) it is enough to show that $y * r' \in \hBR_{\langle a_0, \ldots, a_n \rangle}(q_{\langle a_0, \ldots, a_n \rangle})$. But since $\omega(\langle a_0, \ldots, a_n \rangle^+) \geq n+1$, this indeed follows by the definition of $\hBR$, and the assumptions on $y$ and $r'$.
$\hfill \Box$

%%%%%%%%%%%%%%%%%%%%%%%%%%%%%%%%%%%%%%%%%%%%%%%%%%%%%%%%%%%%%
%  This is the lemma corresponding to the main lemma on Spector's bar recursion (which solves Spector's equstions)      %
%%%%%%%%%%%%%%%%%%%%%%%%%%%%%%%%%%%%%%%%%%%%%%%%%%%%%%%%%%%%%

\begin{lemma} \label{lemma-main-eq-basic} Let $t$ and $a_i$'s be as above. Define $N = 1 + \max \{ |s| \; \colon \; s \in t \}$. For some $n < N$ we have that
\begin{itemize}
	\item[($a$)] $n$ is the least such that $\omega(\langle a_0, \ldots, a_n \rangle^+) < n + 1$, and \\[-2mm]
	\item[($b$)] $\langle a_0, \ldots, a_n \rangle \in t$.
\end{itemize}
\end{lemma}
{\bf Proof}. Suppose that for all $n \leq N$ we have $\omega(\langle a_0, \ldots, a_{n-1} \rangle^+) \geq n$. By Lemma \ref{lemma-pre-eq-basic} this would imply $\langle a_0, \ldots, a_{N-1} \rangle * r \in t$ for some non-empty finite sequence $r$, which is a contradiction by the definition of $N$. Therefore, let $n < N$ be the smallest such that $\omega(\langle a_0, \ldots, a_n \rangle^+) < n + 1$, so that for all $i \leq n$ we have $\omega(\langle a_0, \ldots, a_{i-1} \rangle^+) \geq i$. By Lemma \ref{lemma-pre-eq-basic} again we have that $\langle a_0, \ldots, a_{n-1} \rangle * a_n * r \in t$ for all $r \in \hBR_{\langle a_0, \ldots, a_n \rangle}(q_{\langle a_0, \ldots, a_n \rangle})$. But since $\omega(\langle a_0, \ldots, a_n \rangle^+) < n + 1$ we have that $\hBR_{\langle a_0, \ldots, a_n \rangle}(q_{\langle a_0, \ldots, a_n \rangle}) = \{\emptyseq\}$, implying $\langle a_0, \ldots, a_n \rangle \in t$.
$\hfill \Box$

\begin{lemma} \label{lemma-main-eq} Let $t, p_i, a_i$ be as above, and $n < N$ as in Lemma \ref{lemma-main-eq-basic}. Let also $s = \langle a_0, \ldots, a_n \rangle$. Then for all $i \leq n$ 
\begin{equation}
\begin{array}{lcl}
	q(s) & \subseteq & p_i(a_i).
\end{array}
\end{equation}
\end{lemma}
{\bf Proof}. By Lemma \ref{lemma-main-eq-basic} we have that $s \in t$. Hence, by Lemma \ref{lemma-basic}, for $i \leq n$
\eqleft{s \in \{ \langle a_0, \ldots, a_{i-1}, a_i \rangle * r \; \colon \; r \in \hBR_{\langle a_0, \ldots, a_{i-1}, a_i \rangle}(q_{\langle a_0, \ldots, a_{i-1}, a_i \rangle}) \}.}
It follows that
\eqleft{q(s) \in \{ q(\langle a_0, \ldots, a_{i-1}, a_i \rangle * r) \; \colon \; r \in \hBR_{\langle a_0, \ldots, a_{i-1}, a_i \rangle}(q_{\langle a_0, \ldots, a_{i-1}, a_i \rangle}) \}.}
Hence
\eqleft{
\begin{array}{lcl}
q(s)
	& \subseteq & \bigcup \{ q_{\langle a_0, \ldots, a_{i-1}, a_i \rangle}(r) \; \colon \; r \in \hBR_{\langle a_0, \ldots, a_{i-1}, a_i \rangle}(q_{\langle a_0, \ldots, a_{i-1}, a_i \rangle}) \} \\[2mm]
	& = & p_i(a_i)
\end{array}
}
which concludes the proof.
$\hfill \Box$

%%%%%%%%%%%%%%%%%%%%%%%%%%%%%%%%%%%%%%%%%%%%%%
%%%%%%%%%%%%%%%%%%%%%%%%%%%%%%%%%%%%%%%%%%%%%%
\section{Application: Herbrand Interpretation of $\DNS$}
%%%%%%%%%%%%%%%%%%%%%%%%%%%%%%%%%%%%%%%%%%%%%%
%%%%%%%%%%%%%%%%%%%%%%%%%%%%%%%%%%%%%%%%%%%%%%
\label{sec-herbrand}

In this final section we show how the product of $T$-selection functions, with $T$ being the finite power-set monad, witnesses the Herbrand functional interpretation of the double negation shift
\[ \DNS \quad \colon \quad \forallSt n^\NN \neg \neg A(n) \to \neg \neg \forallSt n^\NN A(n) \]
where $\forallSt x A(x)$ is the quantification over standard objects from \cite{BBS(2012)}. Let us first briefly recall here the definition of the Herbrand functional interpretation from \cite{BBS(2012)}. We shall only present the $\{\to, \forallSt, \bot\}$-fragment as this is enough to carry out the interpretation of $\DNS$. Negation $\neg A$ is defined as $A \to \bot$. Although we will present an explicit definition for the witnesses of $\DNS$, for simplicity, we will carry out the \emph{verification of correctness} in a classical setting, reading the weak existential $\neg \forall x \neg A$ as the strong one $\exists x A$.

\begin{definition}[\cite{BBS(2012)}] The Herbrand functional interpretation of a formula $A$ is defined by structural induction. Assume\footnote{Here $a, b, c$ and $d$ are potentially tuples of variables, though for simplicity we will treat them as if they were single variables.} $\herb{A} = \existsSt a^{X} \forallSt b^Y A_H(a, b)$ and $\herb{B} = \existsSt c^V \forallSt d^W B_H(c, d)$. The only relevant cases for the interpretation of $\DNS$ are:
\[
\begin{array}{lcl}
	\herb{\bot} & \equiv &  \bot \\[2mm]
	\herb{A \to B} & \equiv &  \existsSt f, g \forallSt a^{X}, d^W (\forall b \!\in\! \monAppD{g}{a}{d} \, A_H(a, b) \to B_H(\monApp{f}{a}, d)) \\[2mm]
	\herb{\forallSt z^Z A} & \equiv &  \existsSt h^{Z \Rightarrow X} \forallSt z, b A_H(\monApp{h}{z}, b)
%	\herb{\exists z^Z A} & \equiv &  \exists a^{X} \exists c^{\power{Z}} \forall d^{\power{Y}} \exists z \!\in\! c \forall b \!\in\! d \, A_H(a, b)
\end{array}
\]
where in the clause for $A \to B$ the types of $f$ and $g$ are
\[
\begin{array}{lcl}
f & \colon & X \Rightarrow V \\[2mm]
g & \colon & X \Rightarrow (W \Rightarrow \power{Y}).
\end{array}
\]
For all other cases, including the other base cases, see \cite{BBS(2012)}.
\end{definition}

Let us start by working out the Herbrand interpretation of negation $\neg A$ and double-negation $\neg \neg A$. If $\herb{A} = \existsSt a^{X} \forallSt b^R A_H(a, b)$ then
\eqleft{\herb{\neg A} \equiv \existsSt p^{X \Rightarrow \power{R}} \forallSt a^{X} \neg \forall b \in \monApp{p}{a} A_H(a, b)}
and hence
\eqleft{(\neg\neg A)^H \equiv \existsSt \varepsilon \forallSt p^{X \Rightarrow \power{R}} \exists a^X \!\in \monApp{\varepsilon}{p} \forall b \in \monApp{p}{a} A_H(a, b)}
where $\varepsilon \colon (X \Rightarrow \power{R}) \Rightarrow \power{X}$. Assuming $A(n)$ has a Herbrand functional interpretation $\existsSt a^X \forallSt b^R A_H(n, a, b)$ then the interpretation of $\forallSt n^\NN \neg \neg A(n)$ is
\begin{equation}
\existsSt \delta \forallSt n \forallSt p^{X \Rightarrow \power{R}} \exists a \in \monAppD{\delta}{n}{p} \forall b \in \monApp{p}{a} A_H(n, a, b).
\end{equation}
The interpretation of the conclusion of $\DNS$, $\neg \neg \forallSt n^\NN A(n)$, follows from\footnote{Instead of producing a set of functions $\beta \colon \NN \Rightarrow X$ we will actually produce a single function $\beta \colon \NN \to X$. Note that $\monApp{\{p\}}{a} = p(a)$.}
\begin{equation}
\existsSt \alpha \forallSt \varphi, q \exists \beta \in \monAppD{\alpha}{\varphi}{q} \forall n \in \monApp{\varphi}{\beta} \forall b \in \monApp{q}{\beta} A_H(n, \beta(n), b)
\end{equation}
where the types above are
\eqleft{
\begin{array}{lcl}
\delta \colon \NN \Rightarrow (X \Rightarrow \power{R}) \Rightarrow \power{X} & \quad & p \colon X \Rightarrow \power{R} \\[2mm]
q \colon (\NN \to X) \Rightarrow \power{R} & & \beta \colon \NN \to X \\[2mm]
\varphi \colon (\NN \to X) \Rightarrow \power{\NN} & & \monAppD{\alpha}{\varphi}{q} \colon \power{\NN \Rightarrow X}.
\end{array}
}
Given $\delta, \varphi$ and $q$, we will calculate finite sets $\alpha, N$ and $P$ and show that
\[
\begin{array}{l}
\forall n \!\in\! N \forall p \!\in\! P \exists a \!\in\! \monAppD{\delta}{n}{p} \forall b \!\in\! \monApp{p}{a} A_H(n, a, b) \\[2mm]
\quad \quad \to \exists \beta \!\in\! \alpha  \forall n \!\in\! \monApp{\varphi}{\beta} \forall b \!\in\! \monApp{q}{\beta} A_H(n, \beta(n), b).
\end{array}
\]
Although the Herbrand interpretation here would only actually ask us to produce finite sets of candidate ``constructions" for $\alpha, N$ and $P$, with a guarantee that one of them did the job, we show that in fact we can produce concrete finite sets $\alpha, N$ and $P$.  Given $\delta, \varphi$ and $q$ as above, let us define
\eqleft{ 
\begin{array}{l}
\varepsilon_n \colon (X \to \power{R}) \to \power{X} \\[2mm]
\hat{q} \colon X^* \to \power{R} \\[2mm]
\omega \colon (\NN \to X) \to \NN
\end{array}
}
as
\eqleft{
\begin{array}{lcl}
\varepsilon_n(p) & = & \monAppD{\delta}{n}{\{p\}} \\[2mm]
\hat{q}(s) & = & \monApp{q}{s^+} \\[2mm]
\omega(\beta) & = & \max (\monApp{\varphi}{\beta}).
\end{array}
}
We will then apply $\hBR$ to $\varepsilon_n$, $\hat{q}$ and $\omega$.

\begin{theorem} \label{thm-main} Define $t = \hBR_{\emptyseq}^\omega(\varepsilon)(\hat{q})$. We claim that
\eqleft{
\begin{array}{lcl}
	\alpha & = & \{ s^+ \; \colon \; s \in t \} \\[2mm]
	P & = & \{ p_r \; \colon \; r \preceq s \wedge s \in t \} \\[2mm]
	N & = & 1 + \max \{ |s| \; \colon \; s \in t \}
\end{array}
}
where $p_r(y) = \bigcup \{ \hat{q}(r * y * r') \; \colon \; r' \in \hBR(r * y) \}$,  witness the Herbrand interpretation of $\DNS$, i.e.
\[
\begin{array}{l}
\forall n \!\leq\! N \forall p \!\in\! P \exists a \!\in\! \monAppD{\delta}{n}{p} \forall b \!\in\! \monApp{p}{a} A_H(n, a, b) \\[2mm]
\hspace{2cm} \to \exists \beta \!\in\! \alpha  \forall i \!\in\! \monApp{\varphi}{\beta} \forall b \!\in\! \monApp{q}{\beta} A_H(i, \beta(i), b)
\end{array}
\]
viewing the number $N$ as the finite set $\{0, 1, \ldots, N\}$.
\end{theorem}
{\bf Proof}. Assume
\begin{equation} \label{assumption}
\forall n \leq N \forall p \!\in\! P \exists a \!\in\! \monAppD{\delta}{n}{p} \forall b \!\in\! \monApp{p}{a} A_H(n,a, b).
\end{equation}
By induction on $n$ it follows that: For all $n \leq N$ there exists a sequence $\langle a_0, \ldots, a_n \rangle$ such that either
\begin{itemize}
	\item for some $i < n$, $\omega(\langle a_0, \ldots, a_i \rangle^+) < i + 1$,  or
	\item for all $i \leq n$, 
	\begin{equation} \label{fin-seq}
	a_i \!\in\! \underbrace{\monAppD{\delta}{i}{\{p_{\langle a_0, \ldots, a_{i-1} \rangle}\}}}_{\varepsilon_i(p_{\langle a_0, \ldots, a_{i-1} \rangle})} \,\wedge\; \forall b \!\in\! \underbrace{\monApp{\{p_{\langle a_0, \ldots, a_{i-1} \rangle}\}}{a_i}}_{p_{\langle a_0, \ldots, a_{i-1} \rangle} (a_i)} A_H(i,a_i, b).
	\end{equation}
\end{itemize}
We have used Lemma \ref{lemma-pre-eq-basic}, since under the assumption that $\omega(\langle a_0, \ldots, a_{i-1} \rangle^+) \geq i$ for all $i \leq n$ then $\langle a_0, \ldots, a_{i-1} \rangle * r \in t$, for some $r$, and hence $p_{\langle a_0, \ldots, a_{i-1} \rangle} \in P$. By Lemma \ref{lemma-main-eq-basic} there exists a least $n < N$ such that $\omega(\langle a_0, \ldots, a_n \rangle^+) < n + 1$, so that (\ref{fin-seq}) holds for all $i \leq n$, and $\langle a_0, \ldots, a_n \rangle \in t$. Let $s = \langle a_0, \ldots, a_n \rangle$ and $\beta = s^+$ (so that $\beta \in \alpha$). Note that
\eqleft{\max(\monApp{\varphi}{s^+}) = \omega(s^+)  < |s|.} 
Hence, $i < |s|$ for all $i \in \monApp{\varphi}{s^+}$. By Lemma \ref{lemma-main-eq}
\eqleft{\monApp{q}{\beta} = \monApp{q}{s^+} =  \hat{q}(s) \subseteq p_{\langle a_0, \ldots, a_{i-1} \rangle}(a_i)$, for all $i \in \monApp{\varphi}{s^+}.}
By (\ref{fin-seq}) we can conclude that $\forall i \!\in\! \monApp{\varphi}{\beta} \forall b \!\in\! \monApp{q}{\beta} A_H(i, \beta(i), b)$.
$\hfill \Box$

A reader familiar with the bounded functional interpretation of $\DNS$ (cf. \cite{FE(2010)}) will have noticed several similarities with the Herbrand functional interpretation of $\DNS$ presented here. The main difference, however, is that we have made no effort to formalise the \emph{verification} of the interpretation in a constructive setting, choosing to view $\neg \forall x \neg A$ as a strong existence $\exists x A$. Although it is clear to us that such formalisation is possible, attempting to do so would complicate the verification and probably obfuscate the crucial steps of the bar recursive construction. We hope that by simplifying the ``logical component" of the proof one can better appreciate its ``computational" aspect and the use of the ``Herbrand" bar recursion. The recent paper \cite{Ferreira(2015A)} sheds some light at the relationship between the two interpretations. 

%%%%%%%%%%%%%%%%%%%%%%%%%%%%%%%%%%%%%%%%%%%%%%
%%%%%%%%%%%%%%%%%%%%%%%%%%%%%%%%%%%%%%%%%%%%%%
\section{Conclusion}
%%%%%%%%%%%%%%%%%%%%%%%%%%%%%%%%%%%%%%%%%%%%%%
%%%%%%%%%%%%%%%%%%%%%%%%%%%%%%%%%%%%%%%%%%%%%%

We conclude by noticing that all lemmas of Section \ref{sec-lemmas} were proven for the specific case of the finite power set monad only. It is reasonable to ask whether more general versions of such lemmas work already for the monadic bar recursion $\TEPS$. The main challenge as we see it is to find the appropriate abstraction to the notion of set containment and subset inclusion. Similarly, one might consider generalisations of the Herbrand functional interpretation whereby the finite power set monads is replaced by an arbitrary monad, with possibly some extra structure. 

\bibliographystyle{plain}

\bibliography{../../dblogic.bib}

% \tableofcontents

\end{document}